\documentclass[10pt,twocolumn,english,aps,manuscript]{revtex4}
\usepackage[T1]{fontenc}
\usepackage[latin9]{inputenc}
\usepackage[a4paper]{geometry}
\usepackage{color}
\usepackage{amstext}
\usepackage{graphicx}

\makeatletter
\@ifundefined{textcolor}{}
{%
 \definecolor{BLACK}{gray}{0}
 \definecolor{WHITE}{gray}{1}
 \definecolor{RED}{rgb}{1,0,0}
 \definecolor{GREEN}{rgb}{0,1,0}
 \definecolor{BLUE}{rgb}{0,0,1}
 \definecolor{CYAN}{cmyk}{1,0,0,0}
 \definecolor{MAGENTA}{cmyk}{0,1,0,0}
 \definecolor{YELLOW}{cmyk}{0,0,1,0}
 }

\makeatother

\usepackage{babel}

\begin{document}

\title{Colossal Thermoelectric Power Factor in K$_{7/8}$RhO$_2$}

\author{Y. Saeed, N. Singh, and U. Schwingenschlgl}

\address{PSE Division, KAUST, Thuwal 23955-6900, Kingdom of Saudi Arabia}
\begin{abstract}
We discuss the thermoelectric and optical properties of layered K$_{x}$RhO$_{2}$
(\emph{x} = 1/2 and 7/8) in terms of the electronic structure determined
by first principles calculations as well as Boltzmann transport theory.
Our optimized lattice constants differ significantly from the experiment,
but result in optical and transport properties close to the experiment.
The main contribution to the optical spectra are due to intra and
inter-band transitions between the Rh 4\emph{d} and O 2\emph{p} states.
We find a similar power factor for pristine K$_{x}$RhO$_{2}$ at
low and high cation concentartions. Our transport results of hydrated
K$_{x}$RhO$_{2}$ at room temperature show highest value of the power
factor among the hole-type materials. Specially at 100 K, we obtain
a value of 3$\times$10$^{-3}$ K$^{-1}$ for K$_{7/8}$RhO$_{2}$,
which is larger than that of Na$_{0.88}$CoO$_{2}$ {[}M. Lee \emph{et
al}., Nat. Mater. 5, 537 (2006){]}. In general, the electronic and
optical properties of K$_{x}$RhO$_{2}$ are similar to Na$_{x}$CoO$_{2}$
with enhanced transport properties in the hydrated phase. 


\end{abstract}
\maketitle
Layered cobalt oxides are of technological interest because of their
transport properties \cite{key-1,key-2,key-3,key-4,Nat.Mat.-2006}.
By varying the Na concentration in Na$_{x}$CoO$_{2}$ the system
changes its behaviour from metallic to insulating and becomes superconducting
near \emph{x} = 0.3 \cite{key-3,key-4}. Recently, a strong enhancement
of the Seebeck coefficient has been reported for
Na$_{0.88}$CoO$_{2}$ at T = 80 K \cite{Nat.Mat.-2006}, which greatly
improves the prospects for thermoelectric applications. The peak value
of Seebeck coefficient and the power factor were
found to be 200 $\mu$V/K and 1.8$\times$10$^{-3}$ K$^{-1}$, respectivily,
which is among the highest values for hole-type materials below 100
K. This fact has promoted huge interest in the isostructural and isovalent
families A$_{x}$CoO$_{2}$ (A = K, Rb, Cs). Angle-resolved photoemission
spectroscopy points to similar electronic and optical properties of
K$_{1/2}$CoO$_{2}$ and Na$_{1/2}$CoO$_{2}$, which is also confirmed
by band structure calculations \cite{key-14,key-15,key-16,key-20}.
The electronic structures of hydrated and unhydrated Na$_{x}$CoO$_{2}$
is studied by first principles calculations \cite{singh-h2o}.

Analogous compounds with Rh in place of Co are found to be good thermoelectric
materials with reduced correlation effects \cite{key-20,key-23,key-25,key-26,key-27}.
Shibasaki \emph{et al}. \cite{Rh-2011} have shown that the substitution
of Rh ions in La$_{0.8}$Sr$_{0.2}$Co$_{1-x}$Rh$_{x}$O$_{_{3-\delta}}$
diminishes the magnetic moment of Co, where the thermopower is enhanced
by a factor of 10 at \emph{x} = 1/2 as compared to \emph{x} = 0 and
1. The electronic structure, optical and thermoelectric properties
of K$_{0.49}$RhO$_{2}$ is investigated by Okazaki \emph{et al.}
\cite{optic-2011,krho-40uv/k} and found a qualitative similarities
in the optical conductivity spectra as compared to Na$_{x}$CoO$_{2}$.
The experimental Seebeck coefficient of 40 $\mu$V/K (300 K) is reported
\cite{krho-40uv/k}. The temperature dependence of transport properties
is different from those of Na$_{x}$CoO$_{2}$, which also suggests
that the correlation in the Rh oxides is weaker than in Na$_{x}$CoO$_{2}$.
An enhancement of the transport properties for an increasing concentration
of alkali cations has been reported for other layered oxides as well
\cite{Li-PRB-2011}. Though, various 4\emph{d} systems have been investigated,
the electronic structure in general is poorly understood so far \cite{Axrho-1,Axrho-2}.

In order to throw light on the inter connection between enhancement
of thermoelectric properties and a high cation concentration, first
principles calculations of electronic, optical, and thermoelectric
properties for K$_{x}$RhO$_{2}$ (\emph{x} = 1/2 and 7/8) are performed.
The experimental data of optical and transport properties is taken
from Refs. \cite{krho-40uv/k,optic-2011}. A comparison to Na$_{x}$CoO$_{2}$
is given in terms of the chemical nature of the Co 3\emph{d} and Rh
4\emph{d} orbitals. The optical transitions are explained by the electronic
band structure (BS) and density of states (DOS). The influence of
a high cation concentration on power factor is also discussed, which
is key for thermoelectric devices.

\begin{figure}
\includegraphics[scale=0.3]{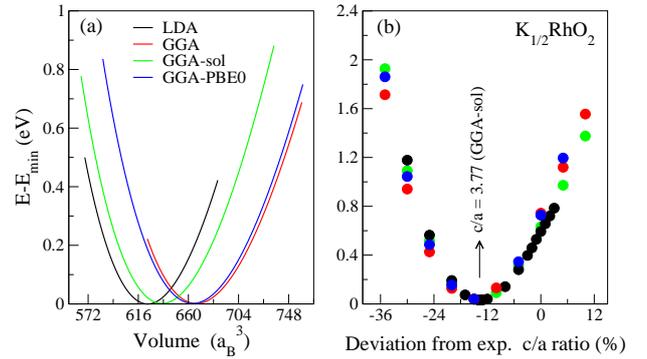}\caption{Volume optimization of K$_{1/2}$RhO$_{2}$ for different exchange
correlation functionals.}

\end{figure}

Our calculations are based on density functional theory (DFT), using
the full-potential linearized augmented plane wave approach as implemented
in the WIEN2k code \cite{wien2k}. This method has been successfully
applied to describe the electronic structure of oxides \cite{U1,U2}
including the optical spectrum \cite{N.Singh}. The transport is calculated
by the semiclassical Boltzmann theory within the constant scattering
approximation, as implemented in the BoltzTraP code \cite{BoltzTraP}.
Various exchange and correlation functionals (local density approximation
(LDA), generalized gradient approximation (GGA), GGA-sol, and GGA-PBE0)
are used for optimizing the \emph{c/a} ratio. Since the differences
are small, we will discuss in following only GGA-sol results for the
electronic, optical and transport properties. 

The unit cell is divided into non-overlapping atomic spheres centered
at the atomic sites and an interstitial region. The convergence parameter
R$_{mt}$K$_{max}$, where K$_{max}$ is the plane-wave cut-off and
R$_{mt}$ is the smallest of all muffin-tin radii controls the size
of the basis set. This convergance parameter is set to 7 together
with G$_{max}$=24. We use 66 \textbf{k}-points in the irreducible
wedge of the Brillouin zone for calculating the electronic structure
and a dense mesh of 480 \textbf{k}-points in the optical calculations.
For the transport calculations, we use 4592 \textbf{k}-points. Self-consistency
is assumed to be achieved for a total energy convergence of $10^{-5}$
Ryd. 

\begin{figure}
\includegraphics[scale=0.2]{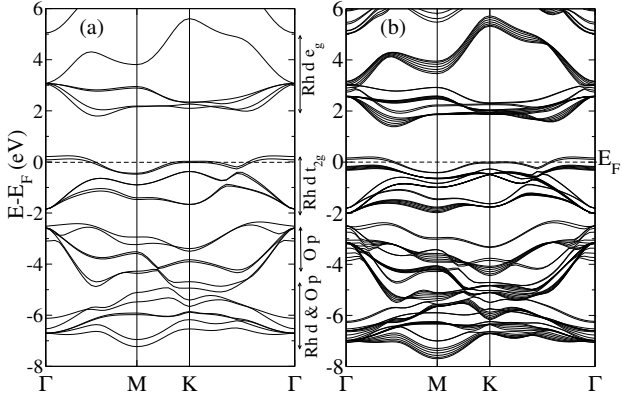}\caption{Energy band structures of (a) K$_{1/2}$RhO$_{2}$ and (b) K$_{7/8}$RhO$_{2}$. }

\end{figure}

K$_{x}$RhO$_{2}$ crystallizes in the $\gamma$-Na$_{x}$CoO$_{2}$
structure (space group P6$_{\text{3}}$/mmc) with the experimental
lattice constants \emph{a} = 3.0647  and \emph{c} = 13.6  \cite{yubuta-exp a/c}.
The CdI$_{\text{2}}$-type RhO$_{\text{2}}$ layer and the K layer
are stacked alternately along the \emph{c}-axis. The experimental
lattice constants of K$_{1/2}$RhO$_{2}$ are used as a starting point
of the optimization, obtaining the optimized volume and \emph{c/a}
ratio presented in Fig. 1. Our LDA calculation yields a $\sim$ 14\%
reduction of the \emph{c/a} ratio from 4.44 (experimental) to 3.84
(calculated), which is close to other layered Co/Rh oxides (\emph{c/a}
ratio $\sim$ 3.8). We obtain lattice constants of\emph{ a} = 3.02
 and \emph{c} = 11.63 . The bonding length between Rh and O is reduced
from 2.1326  to 2.0395 . To confirm the large deviation of the\emph{
c/a} ratio from the experimental structure parameters, we have optimized
the\emph{ }structure\emph{ }by more involved exchange correlation
functionals (GGA, GGA-sol and PBE0). However, we obtain almost the
same results for all functionals. In addition, our optimized lattice
parameters are confirmed by calculations of the optical and transport
properties, see the discussion below. The calculated Seebeck coefficient
is overestimated for the experimental lattice constants, while the
optimized lattice constants yield a Seebeck coefficient close to the
experiment. The calculated optical conductivity of K$_{1/2}$RhO$_{2}$
(with optimized lattice constants) at zero photon energy is found
to be $\sim$ 2500 $\Omega^{-1}$cm$^{-1}$,which is agrees excellently
with the experiment.

The observed difference of the \emph{c} length between K$_{x}$RhO$_{2}$
and Na$_{x}$CoO$_{2}$ could be attributed to the different ionic
radii of K and Na. However, our optimized value of \emph{c} is similar
to other isostructural layered oxides such as Sr$_{x}$RhO$_{2}$
\cite{SrRhO2,sro2006}, Na$_{x}$CoO$_{2}$ \cite{key-15}, Li$_{x}$NbO$_{2}$
\cite{LiNbO2}, and La$_{x}$CoO$_{2}$ \cite{LaCoO2}. In addition,
LiRhO$_{2}$, NaRhO$_{2}$, and KRhO$_{2}$ layered oxides can form
a hydrate (water intercalated) phase \cite{Axrho-1} with an increased
\emph{c} length \cite{Park}. Takada \emph{et al.} \cite{key-4} showed
that Na$_{x}$CoO$_{2}$ can be readily hydrated to form Na$_{x}$CoO$_{2}$
\emph{y}H$_{\text{2}}$O, maintaining the CoO$_{\text{2}}$ layers,
but with a considerably expanded \emph{c} axis, to accomodate the
intercalated water.  In the following, the optimized lattice constants
are used, unless stated otherwise.

\begin{figure}
\includegraphics[scale=0.35]{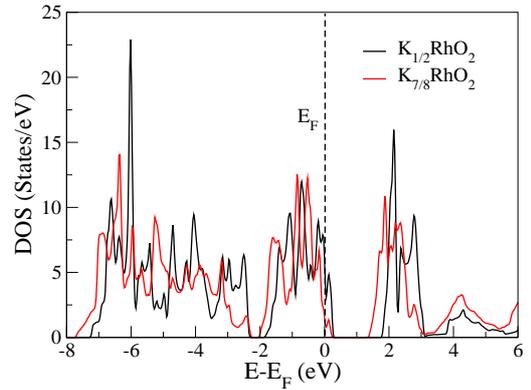}\caption{DOS obtained for K$_{1/2}$RhO$_{2}$ and K$_{7/8}$RhO$_{2}$ .}

\end{figure}

In Figs. 2 (a) and (b) the calculated electronic BSs of K$_{1/2}$RhO$_{2}$
and K$_{7/8}$RhO$_{2}$is presented. The BS of K$_{1/2}$RhO$_{2}$
is similar to isostructural and isovalent Na$_{1/2}$CoO$_{2}$, except
for a slightly larger pseudogap. Moreover, the strongly dispersive
bands and high hole concentration in K$_{7/8}$RhO$_{2}$ give rise
to a high thermoelectricity. The calculated DOS (Fig. 3) shows a crystal
field splitting experienced by the Rh$^{4+}$ ions (splitting into
$e_{g}$ and $t_{2g}$ states) is similar to the Co$^{3+}$ case,
but with a larger bandwidth. The bandwidth of the $t_{2g}$ state
is 1.52 eV for Rh$^{4+}$ and 1.46 eV for Co$^{3+}$ in the case of
K$_{1/2}$CoO$_{2}$ (not shown here). A similar increment is also
observed for the $e_{g}$ states, in agreement with the experiment
\cite{optic-2011}. In addition, a weak hybridization is observed
between the Rh 4\emph{d} and O 2\emph{p} states at/below the Fermi
level. The O\emph{ p} states lie deep in the valence band (below $-$2
eV) as compared to Na$_{0.50}$CoO$_{2}$.\textcolor{red}{{} }The DOS
of K$_{7/8}$RhO$_{2}$ shows increased bandwidths of the $e_{g}$
and $t_{2g}$ states, which reflects a reduction of the electronic
correlation effects in K$_{7/8}$RhO$_{2}$ as compared to Na$_{x}$CoO$_{2}$. 

\begin{figure}
\includegraphics[scale=0.35]{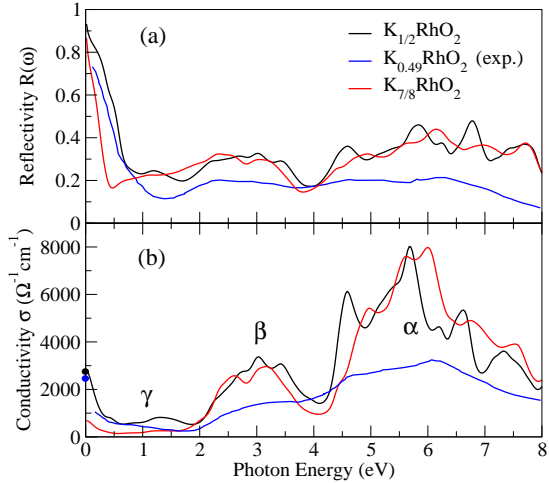}\caption{Optical reflectivity and conductivity of K$_{1/2}$RhO$_{2}$ and
K$_{7/8}$RhO$_{2}$ along with experimental data \cite{optic-2011}.
Blue and black dots respresent the calculated and experimental values
at zero photon energy. }

\end{figure}

The optical properties of K$_{x}$RhO$_{2}$ (\emph{x} = 1/2 and 7/8)
are studied and presented in Figs. 4(a) and 4(b). The experimental
results are taken from Ref. {[}17{]}. The obtained reflectivities
of K$_{1/2}$RhO$_{2}$ and K$_{7/8}$RhO$_{2}$ (Fig. 4(a)) are similar
to each other with a maximum value of $\sim$ 90\% near 0 eV. A Drude-like
edge in the optical reflectivity of K$_{1/2}$RhO$_{2}$ is found
at $\sim$ 1 eV (experiment: 1.2 eV), while for K$_{7/8}$RhO$_{2}$
this edge appears at $\sim$ 0.5 eV. At zero photon energy, the calculated
optical conductivity of $\sigma\sim$ 2500 $\Omega^{-1}$cm$^{-1}$
for K$_{1/2}$RhO$_{2}$ is obtained in excellent agreement with the
experiment (blue and black dots in fig. 4(b)). Three well defined
peaks are observed: (i) near 1 eV due to the intra-band transition
of Rh ($t_{2g}$-$t_{2g}$), (ii) at $\sim$ 3 eV due to the inter-band
transition of Rh 4\emph{d} ($t_{2g}$-$e_{g}$), and (iii) around
5.5 eV due to the inter-band transition from the O 2\emph{p} to the
Rh 4\emph{d} $e_{g}$ states. These peaks are also present in the
experiment \cite{optic-2011} as well as for Na$_{1/2}$CoO$_{2}$
(0.5 eV, 1.6 eV, and 3 eV, respectively) \cite{key-14}, which again
reflects the similarity between these isostructural and isovalent
compounds. 

In the following, we will address both the experimental crystal structure
(hydrated) and optimized structure of K$_{x}$RhO$_{2}$. We have
calculated the Seebeck coefficient (S), thermal conductivity ($\kappa$),
and power factor (Z). The results are plotted in Figs. 5(a), (b),
and (c) as a function of the temperature from 0 to 700 K, and compared
with the experimental Z of Na$_{0.88}$CoO$_{2}$ \cite{Nat.Mat.-2006}.
Fig. 5(a) shows that the calculated S of pristine K$_{x}$RhO$_{2}$
is $\sim$ 50 $\mu$V/K at 300 K, in agreement with the experiment
(40 $\mu$V/K) \cite{krho-40uv/k}. The calculated S values of hydrated
K$_{1/2}$RhO$_{2}$ and K$_{7/8}$RhO$_{2}$ are strongly enhanced,
amounting to $\sim$ 100 $\mu$V/K and $\sim$ 140 $\mu$V/K, respectively.
The calculated S of pristine K$_{x}$RhO$_{2}$ hardly depens on the
K concentration upto 300 K, while for higher temperature for K$_{7/8}$RhO$_{2}$
increases stronger to reach a value of 80 $\mu$V/K at 700 K. In contrast
to this behavior, the calculated S of hydrated K$_{x}$RhO$_{2}$
remains almost constant above 300 K. According to Fig. 5(b) the thermal
conductivity is similar for the hydrated compounds and for pristine
K$_{7/8}$RhO$_{2}$, while its much enhanced for pristine K$_{1/2}$RhO$_{2}$. 

\begin{figure}
\includegraphics[scale=0.4]{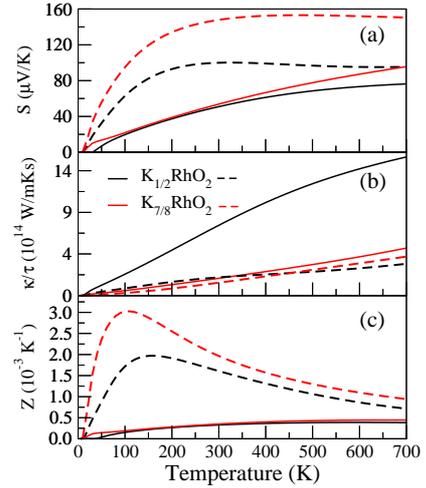}\caption{Calculated thermoelectric properties of prestine (solid line) and
hydrated (dashed line) K$_{x}$RhO$_{2}$. Experimental data from
Ref. \cite{Nat.Mat.-2006} is included.}

\end{figure}

In Fig. 5(c), the calculated power factor Z of pristine and hydrated
K$_{x}$RhO$_{2}$ along with experimental data for Na$_{0.88}$CoO$_{2}$
\cite{Nat.Mat.-2006} is presented. Upto 25 K, power factor of hydrated
K$_{7/8}$RhO$_{2}$ behaves similar to the experimental curve, but
approaches a value of 3$\times$10$^{-3}$ K$^{-1}$ at 100 K, which
is much higher than that of other hole-type materials (like Na$_{0.88}$CoO$_{2}$
Z = 1.8$\times$10$^{-3}$ K$^{-1}$ at 80 K) in this temperature
range. Even at room temperature (300 K), the calculated Z value for
hydrated K$_{x}$RhO$_{2}$ is much higher than Na$_{0.88}$CoO$_{2}$,
while for pristine K$_{x}$RhO$_{2}$ is just slightly greater. The
large power factor in hydrated K$_{x}$RhO$_{2}$ results from a decrease
of the thermal conductivity, increase in the electrical conductivity
as reported for the hydrated phase of NaRhO$_{\text{2}}$ (Fig. 2
of Ref. \cite{Park}), and larger S as compared to that of pristine
K$_{x}$RhO$_{2}$. Therefore, the transport properties of hydrated
K$_{x}$RhO$_{2}$ are highly promising for technological applications.

\begin{figure}

\includegraphics[scale=0.2]{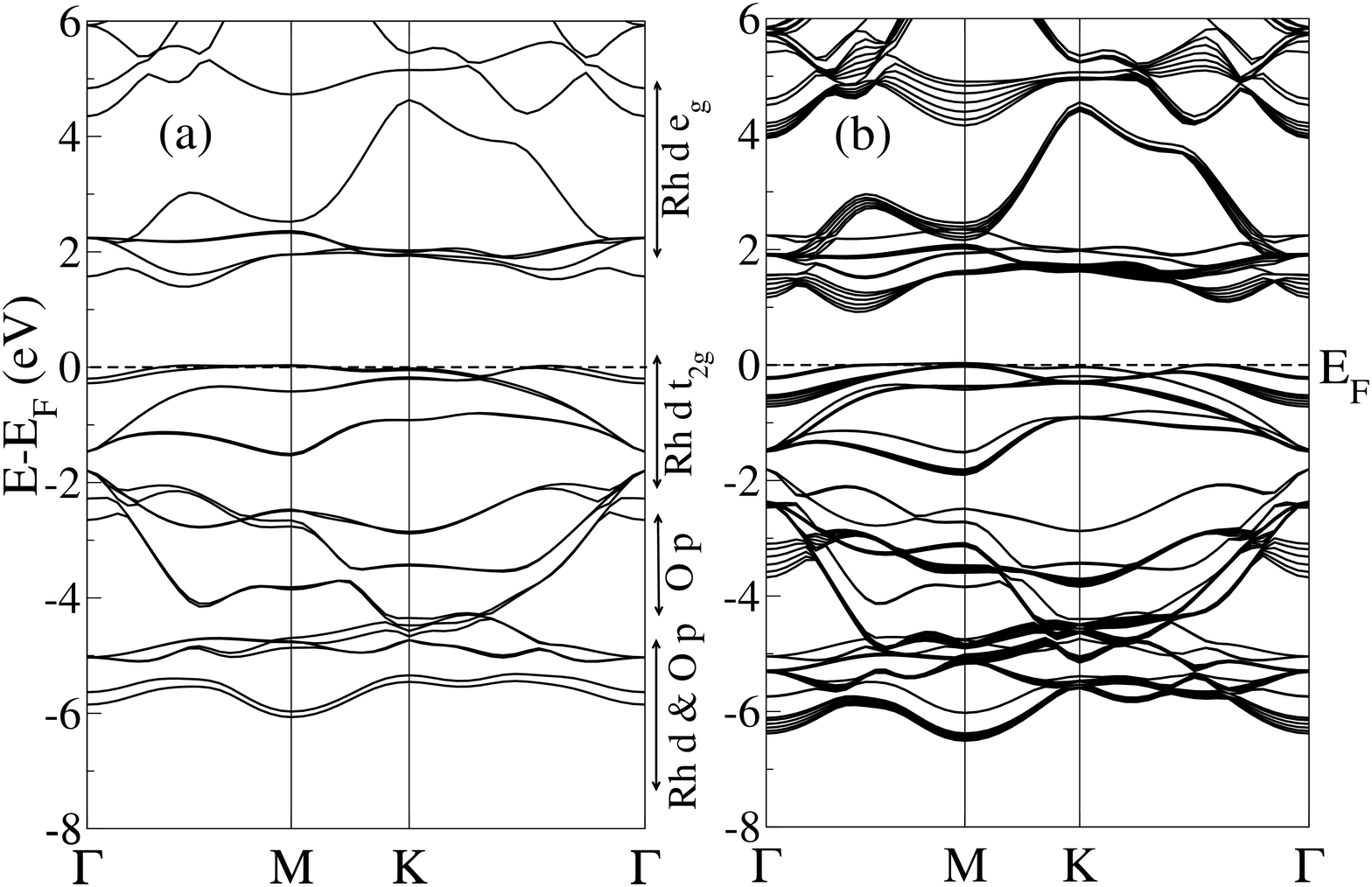}\caption{KRhO2-hydrated}

\end{figure}

In conclusion, the electronic, optical, and transport properties of
layered K$_{x}$RhO$_{2}$ (\emph{x} = 1/2 and 7/8) are calculated,
and compared to the isostructural and isovalent compound Na$_{x}$CoO$_{2}$.
Our optimized structure of K$_{1/2}$RhO$_{2}$ shows a huge deviation
in the \emph{c/a} ratio from the experiment but gives a good agreement
for the optical and transport properties. The large deviations, also
in comparison the other related compounds, indicate that the experimental
structure has been determined for the hydrated phase of K$_{x}$RhO$_{2}$.
The Rh$^{4+}$4\emph{d }$e_{g}$ and $t_{2g}$ states of K$_{1/2}$RhO$_{2}$
are broader than the respective Co$^{3+}$ states in Na$_{x}$CoO$_{2}$,
which confirms previous reports. The calculated Seebeck coefficient
of pristine K$_{x}$RhO$_{2}$ (\emph{x} = 1/2 and 7/8) is S $\sim$
50 $\mu$V/K at room temperature, which is close to the experimental
value of S = 40 $\mu$V/K. Our calculations also show large values
of S and Z for hydrated K$_{x}$RhO$_{2}$ in whole temperature range
from 0 to 700 K. At around 100 K, the calculated Z of hydrated K$_{7/8}$RhO$_{2}$
is 3$\times$10$^{-3}$ K$^{-1}$, which is the highest value in any
hole-type material at this temperature. At room temperature, the calculated
Z value of pristine K$_{x}$RhO$_{2}$ is $\sim$ 0.4$\times$10$^{-3}$
K$^{-1}$, which is also slightly greter than that of Na$_{0.88}$CoO$_{2}$.
Therefore, our results suggest that hydration can be used to elongate
the structure and inducing the doping by the formation of hydronium
ions, results strong enhancement of thermoelectric properties of this
class of layered oxides. 

\begin{figure}
\includegraphics[scale=0.25]{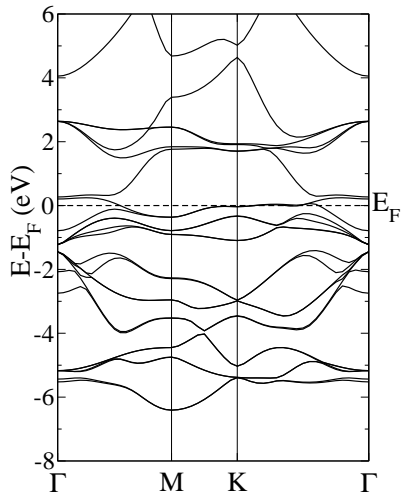}\caption{Band Na50CoO2 c15.6A}

\end{figure}


\begin{thebibliography}{31}
\bibitem{key-1} I. Terasaki, Y. Sasago, and K. Uchinokura, Phys.
Rev. B 56, R12685 (1997).

\bibitem{key-2} T. Motohashi, R. Ueda, E. Naujalis, T. Tojo, I. Terasaki,
T. Atake, M. Karppinen, and H. Yamauchi, Phys. Rev. B 67, 064406 (2003).

\bibitem{key-3} M. L. Foo, Y. Wang, S. Watauchi, H. W. Zandbergen,
T. He, R. J. Cava, and N. P. Ong, Phys. Rev. Lett. 92, 247001 (2004).

\bibitem{key-4} K. Takada, H. Sakurai, E. Takayama-Muromachi, F.
Izumi, R. A. Dilanian, and T. Sasaki, Nature (London) 422, 53 (2003).

\bibitem{Nat.Mat.-2006} M. Lee, L. Viciu, L. Li, Y. Wang, M. L. Foo,
S. Watauchi, R. A. Pascal, R. J. Cava, and N. P. Ong, Nat. Mater.
5, 537 (2006).

\bibitem{key-14} M. D. Johannes, I. I. Mazin, and D. J. Singh, Phys.
Rev. B 71, 205103 (2005).

\bibitem{key-15} J. Sugiyama, H. Nozaki, Y. Ikedo, K. Mukai, J. H.
Brewer, E. J. Ansaldo, G. D. Morris, D. Andreica, A. Amato, T. Fujii,
and A. Asamitsu, Phys. Rev. Lett. 96, 037206 (2006). 

\bibitem{key-16} J. Sugiyama, Y. Ikedo, P. L. Russo, H. Nozaki, K.
Mukai, D. Andreica, A. Amato, M. Blangero, and C. Delmas, Phys. Rev.
B 76, 104412 (2007).

\bibitem{key-20} K. W. Lee and W. E. Pickett, Phys. Rev. B 76, 134510
(2007).

\bibitem{singh-h2o}M. D. Johannes and D. J. Singh, Phys. Rev. B 70,
014507 (2004).

\bibitem{key-23} Y. Klein, S. H\'{e}bert, D. Pelloquin, V. Hardy,
and A. Maignan, Phys. Rev. B 73, 165121 (2006).

\bibitem{key-25} S. Shibasaki, W. Kobayashi, and I. Terasaki, Phys.
Rev. B 74, 235110 (2006).

\bibitem{key-26} W. Kobayashi, S. H\'{e}bert, D. Pelloquin, O. Perez,
and A. Maignan, Phys. Rev. B 76, 245102 (2007).

\bibitem{key-27}W. Koshibae, K. Tsutsui, and S. Maekawa, Phys. Rev.
B 62, 6869 (2000).

\bibitem{Rh-2011}S. Shibasaki, I. Terasaki, E. Nishibori, H. Sawa,
J. Lybeck, H. Yamauchi, and M. Karppinen, Phys. Rev. B 83, 094405
(2011).

\bibitem{optic-2011}R. Okazaki, Y. Nishina, Y. Yasui, S. Shibasaki,
and I. Terasaki, Phys. Rev. B 84, 075110 (2011).

\bibitem{krho-40uv/k}S. Shibasaki, T. Nakano, I. Terasaki, K. Yubuta,
and T. Kajitani, J. Phys. Condens. Matter 22, 115603 (2010).

\bibitem{Li-PRB-2011}T. Motohashi, Y. Sugimoto, Y. Masubuchi, T.
Sasagawa, W. Koshibae, T. Tohyama, H. Yamauchi, and S. Kikkawa, Phys.
Rev. B 83, 195128 (2011).

\bibitem{Axrho-1} A. Mendiboure, H. Eickenbusch, and R. Schollhorn,
J. Solid State Chem. 71, 19 (1987).

\bibitem{Axrho-2} A. Varela, M. Parras, and J. M. Gonz\'{a}lez-Calbet,
Eur. J. Inorg. Chem. 2005, 4410 (2005).

\bibitem{wien2k}P. Blaha, K. Schwarz, G. Madsen, D. Kvasicka, and
J. Luitz, WIEN2k, An Augmented Plane Wave + Local Orbitals Program
for Calculating Crystal Properties (TU Vienna, Vienna, 2001).

\bibitem{U1}U. Schwingenschl\"{o}gl and C. Schuster, Phys. Rev.
Lett. 99, 237206 (2007); EPL 79, 27003 (2007). 

\bibitem{U2}U. Schwingenschl\"{o}gl, C. Schuster, and R. Fr\'{e}sard,
EPL 88, 67008 (2009); EPL 81, 27002 (2008).

\bibitem{N.Singh}N. Singh and U. Schwingenschl\"{o}gl, Chem. Phys.
Lett. 508, 29 (2011); N. Singh, S. M. Saini, T. Nautiyal, and S. Auluck,
J. Appl. Phys. 100, 083525 (2006). 

\bibitem{BoltzTraP}\textcolor{black}{G. K. H. Madsen, K. Schwarz,
P. Blaha, and D. J. Singh, Phys. Rev. B 68, 125212 (2003).}

\bibitem{yubuta-exp a/c} K. Yubuta, S. Shibasaki, I. Terasaki, and
T. Kajitani, Philos. Mag. 89, 2813 (2009).

\bibitem{SrRhO2} A. L. Hector, W. Levason, and M. T. Weller, Eur.
J. Solid State Inorg. Chem. 35, 679 (1998). 

\bibitem{sro2006} Y. Okamoto, M. Nohara, F. Akai, and H. Takagi,
J. Phys. Soc. Jap. 75, 023704 (2006).

\bibitem{LiNbO2}K.-W. Lee, J. Kune\v{s}, R. T. Scalettar, and W.
E. Pickett, Phys. Rev. B 76, 144513 (2007).

\bibitem{LaCoO2}K. Kn\'{i}\v{z}ek, J. Hejtm\'{a}nek, M. Mary\v{s}ko,
E. \v{S}antav\'{a}, Z. Jir\'{a}k, J. Bur\v{s}\'{i}k, K. Kirakci,
P. Beran, J. Solid State Chem. 184, 2231 (2011).

\bibitem{Park}S. Park, K. Kang, W. Si, W. S. Yoon, Y. Lee, A. R.
Moodenbaugh, L. H. Lewis, and T. Vogt, Solid State Commun. 135, 51
(2005).
\end{thebibliography}
\end{document}